\documentclass[sigplan,screen,nonacm=true]{acmart}
\usepackage[T1]{fontenc}
\usepackage{algorithm}
\usepackage{algorithmic}
\usepackage{bm,calrsfs}

		\setcopyright{none}
\DeclareMathOperator*{\argmax}{argmax}

\newcommand{\mE}{\mathbb{E}}
\newcommand{\mR}{\mathbb{R}}

\begin{document}

\title{Low Volatility Stock Portfolio \\ Through High Dimensional Bayesian Cointegration}
\author{Parley R Yang}
\email{ry266@cam.ac.uk}
\affiliation{%
	\institution{Faculty of Mathematics, University of Cambridge}
	\city{Cambridge}
	\country{UK}
}

\author{Alexander Y Shestopaloff}
\affiliation{%
	\institution{School of Mathematical Sciences, Queen Mary University of London}
	\city{London}
	\country{UK}
}

\begin{abstract}
We employ a Bayesian modelling technique for high dimensional cointegration estimation to construct low volatility portfolios from a large number of stocks. The proposed Bayesian framework effectively identifies sparse and important cointegration relationships amongst large baskets of stocks across various asset spaces, resulting in portfolios with reduced volatility. Such cointegration relationships persist well over the out-of-sample testing time, providing practical benefits in portfolio construction and optimization. Further studies on drawdown and volatility minimization also highlight the benefits of including cointegrated portfolios as risk management instruments.
\end{abstract}

\maketitle


\section{Introduction}

\subsection{Motivation}\label{MLTF}
Stationarity of a time series is defined as its first moment and auto-covariance being finite and invariant across time. Stock prices typically do not satisfy this assumption, though their lag-difference or log lag-difference does (such as in models assuming Brownian or Geometric Brownian motion). Cointegration, in this context, concerns with finding portfolios of stocks such that the portfolios can be stationary. This is beneficial for volatility reduction, as well as other portfolio applications such as drawdown and Sharpe ratio improvement. 

More technically, let $Y(i,t)$ denote the price of stock $i$ at the end of day $t$. A vector of $p$ stock prices at $t$ is denoted as $Y(\cdot, t)\in \mR^p$, abbreviated as $Y_t$, while a time series over $T$ days of a particular stock $i$ is denoted as $Y(i,\cdot) \in \mR^T$. We write $Y  \sim I(1)$ if it is non-stationary, but the time series $\Delta Y$ defined by $\Delta Y(i,t) := Y(i,t)-Y(i,t-1) \ \forall i,t$ is stationary, i.e. $\Delta Y \sim I(0)$. 

Now, cointegration in the setting of portfolio management concerns finding the dimension $r$ and linearly independent vectors $\alpha_1, ..., \alpha_r $ such that $\alpha_i Y \sim I(0) \ \forall i\in[r]$. High dimensionality acts as a barrier to classical methodologies as $p$ can be large, coupled with a reasonably sized $r$ to be determined. In most cases where $r<p$, a random non-zero vector $\theta \in \mR^p$ leading to the combination $\theta Y$ would stay non-stationary due to $\theta$ not being in the span of the cointegration vectors $\alpha_1,...,\alpha_r$. 

\begin{figure}[t]
	\centering
	\includegraphics[width=\linewidth]{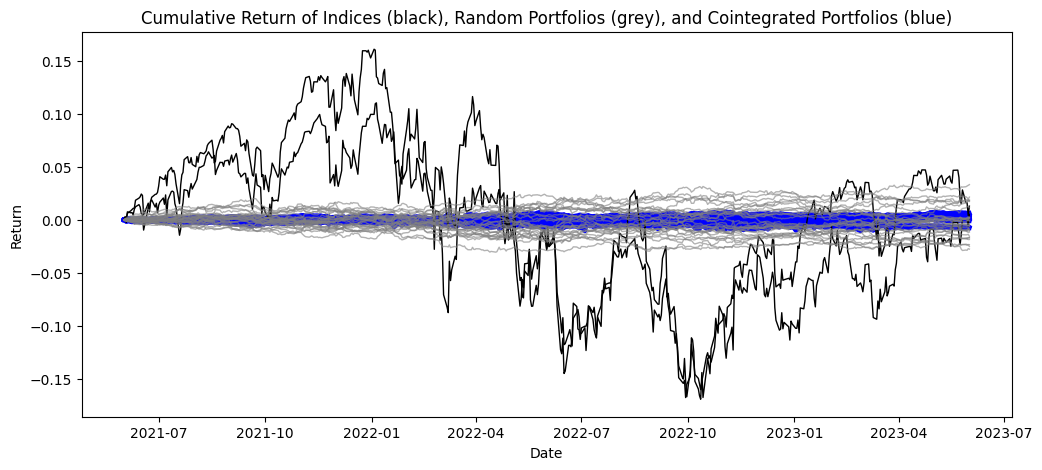}
	\caption{Index (black), random portfolios (grey), and cointegrated portfolio returns}
	\label{fig:1a}
\end{figure}

Additionally, high dimensionality is a clear motive for new models, and has direct implications to portfolio management, where feature dimension can be considered as the number of assets in an investable asset space. For instance, S\&P 500 index has 500 constituents stocks, with more than 400 of them being $I(1)$. If we were to run a Vector Autoregressive model (VAR), even with just 1 lag, the desired main estimation matrix would consist of $p^2$ undetermined parameters --- in daily time series, if we were to match $T$ to $p^2$ this would equate to about 992 years of data.\footnote{Assuming $p=500$ and the daily time series gives 252 observations per year.} In this paper, we utilise a Bayesian model for sparsity to assist modelling high number of features with a low number of observations. This brings down the number of observations required to 2 years, while maintaining the capability to detect the underlining cointegration structure amongst circa 400 to 1000 stocks.

In \autoref{fig:1a}, we illustrate a typical scenario where stock indices and randomly generated unified long-short portfolios (randomly drawn $\theta$ where $||\theta||_1=1$) are plotted in black and grey. Cointegrated portfolios generated by the estimated $\alpha_1, ..., \alpha_r$ using our Bayesian Models after unification\footnote{Unification in the sense that we replace $\alpha_i$ by $\frac{\alpha_i}{||\alpha_i||_1}$} are plotted in blue. The return of cointegrated portfolio in general has a clear low-volatility and a behaviour of more concentration around zero compared to the others, underpinning the topic of this paper's discussion.

\subsection{Literature Review}
There are many existing literature in the fields we fit in, though the combination of them are rare. To start with, portfolio constructions and the relevant statistical studies have been covered in recent literature \cite{BFB22, ZMI22, PL22, TC23}. In particular, Borragiero et al have introduced a ranking algorithm dealing with non-stationarity of stock data \cite{BFB22}. We face the same challenge in the empirics, however, the methodology we proposed relate more intimately to the notion of finding cointegration vectors amongst the non-stationary stock prices. In our application, we engage with the portfolio combination and volatility reduction, which are discussed under various Machine Learning framework such as in \cite{ZMI22, TC23}. Our application on the optimisation front inherits from the Markowitz portfolio \cite{Markowitz}, which have been reviewed extensively in conjunction with penalisation and other modern statistical techniques \cite{PL22, Brodie09}. There are another stream of literature on pairs trading \cite{ElliottHoekMalcolm05}, however, the cointegrated portfolio investigated here rests upon a higher dimensional search of assets --- instead of finding two or three stocks on a pre-selected sub-basket.

In contrast to most of the above literature, the crux of our methodology focus heavily on the cointegration model and Bayesian posterior estimation. To pursue the technical review of literature, we start with the Vector Error Correction Model (VECM) model from a VAR(1), a simplification from \cite{EG1987}:
\begin{align}
	\Delta Y_t &= \Pi Y_{t-1} + \varepsilon_t, \ \ \varepsilon_t\sim {\mathcal N} (0, \Sigma) \label{VAR1} \\
	r & = rank(\Pi)
\end{align}
The rank $r$ leads to the number of unknown cointegrated relationships, paving paths for constructing cointegrated portfolios in stock selection. As $\Pi \in \mR^{p \times p}$ and that we face insufficient data for the estimation, we seek alternative ways to construct an estimator for $\Pi$ and $r$. Liang and Schienle propose the penalised likelihood approach to regulate the sparsity and henceforth $r$ estimation \cite{LiangSchienle19}. Yang and Shestopaloff inherit their partial least square procedure in transforming a pre-estimation of $\Pi$ into an orthogonal decomposition, followed by Bayesian model on the sparse decomposition 	\cite{Yang2023}. Further technical equipments for modelling include Spike-and-Slab Lasso (SSL), which are first presented by Rockova \cite{RG2018}.

\subsection{Contributions}
There are two key contributions we make to both the academic audience and financial practitioners. 

We summarise and extend the theoretical and algorithmic approach in finding high dimensional cointegration relationship (\autoref{Method}) to the practical world, and find evidence of such cointegration in a rolling training and testing manner (\autoref{SPC}). Such cointegration relationships persist well over the out-of-sample testing time, providing practical benefits in portfolio construction and optimisation.

The portfolio optimisation brings benefit to practitioners' risk management, as low volatility portfolio may be engaged in practitioners' strategy combination --- here we assume a simple passive index strategy as the benchmark and found combination with the cointegrated portfolio yield better Sharpe ratio out-of-sample (\autoref{SPcombo}). Further studies are conducted on drawdown and volatility minimisation, which also adds benefit for the inclusion of cointegrated portfolios as instruments for risk management (\autoref{RM}).

\subsection{Notations}
Common notations on distributions are as follows: $ {\mathcal N}$ indicates Gaussian distribution, ${\mathcal B}$ indicates Beta distribution, and $ {\mathcal U}$ indicates uniform distribution. For a vector $\theta \in \mR^p$ and $a<b<p$, $\theta_{a:b} \in \mR^{b-a+1}$ is the vector of entries $a$ to $b$ of $\theta$. For a matrix $R \in \mR^{p\times p} $, the $j$-th column is denoted as $	R(\cdot,j) \in \mR^{p\times 1}$. The set of positive integers up to $p$ is denoted as $[p]:=\{1,2,...,p\}$. Other notations are introduced when they are mentioned.

\section{High Dimensional Bayesian Cointegration Model}\label{Method}

\subsection{Decomposed VECM}

The decomposed VECM could be summarised as follows. Let $Y_t \in \mR^p$ be the multivariate time series of interests. Denote the concatenated matrices $A:= [\Delta Y_1, ..., \Delta Y_T] \in \mR^{p\times T}$ and $B:= [Y_0,.., Y_{T-1}] \in \mR^{p\times T}$, we implement the  model as per \autoref{VAR1} and the following partial least square pre-estimation: \begin{align}
	\tilde{\Pi} &= (AB^\intercal)(BB^\intercal)^{-1} \label{VAR2} 
\end{align}
\autoref{VAR1} is a typical VAR(1) model for the time series, whereas \autoref{VAR2} implements a pre-estimation on the unknown ${\Pi} \in \mR^{p\times p}$. Furthermore, we decompose $\tilde{\Pi}$ into \begin{equation}
\label{decomposition} 	\tilde{\Pi} = \tilde{R}^\intercal \tilde{S}^\intercal 
\end{equation}
where $\tilde{R}$ is a p-dimensional upper triangular matrix and $\tilde{S}$ is an orthogonal matrix. Due to the orthogonality, $rank(\tilde{R}) = rank(\tilde{\Pi}) $ and more crucially, the rank would be the same as the number of non-zero rows of $\tilde{R}^\intercal$. Now, for an alternative estimate $\hat{R}$, we may construct estimator \begin{equation}
\hat{\Pi} = \hat{R}^\intercal \tilde{S}^\intercal  \label{hatPi}
\end{equation}
and henceforth $\hat{r}:=rank(\hat{\Pi}) = rank(\hat{R})$.

The SSL distribution on decomposed VECM is, in this pursuance, tasked to use Bayesian method to detect the sparsity of matrix $R$, for the purpose of estimating the true low rank.

\subsection{SSL Distribution}

We first introduce the Bayesian model behind the high dimensional VECM modelling --- SSL distribution. 	

Let $X\in\mR$ be a random variable. A Spike-and-Slab Lasso (SSL) prior distribution on $X$ is defined as 
\begin{align}
\pi(X|\gamma) :=& (1-\gamma)\psi_0(X) + \gamma \psi_1(X) , \ \ \gamma \in [0,1] \\
\psi_j(x) :=& \frac{\lambda_j}{2}\exp(-\lambda_j |x|) \ \ \forall j \in \{0,1\} \text{ with } \lambda_0> \lambda_1 >0
\end{align}
We write $X|\gamma \sim SSL(\gamma, \lambda_0, \lambda_1)$. Let $X\in\mR^p$ be a random variable. We write $X|\gamma \sim SSL(\gamma, \lambda_0, \lambda_1)$ if for all entries $j \in [p]$, $X_j|\gamma \sim SSL(\gamma, \lambda_0, \lambda_1) \ iid $

For intuition, $SSL(\gamma, \lambda_0, \lambda_1)$ can be thought of as a mixing of two Laplace distributions, with $\psi_0$ being spiky towards zero, while $\psi_1$ being flatter. 

\subsection{SSL Distribution on Decomposed VECM}
We first re-write the \autoref{VAR1} and \autoref{hatPi} with a simplification on $\Sigma = diag(\sigma_1^2,...,\sigma_p^2) $:  $\forall t\in[T], j\in[p],$ \begin{equation}\label{Gaussian}
\Delta Y_{t,j} |Y_{t-1}, R, \sigma_j \sim {\mathcal N} (R(\cdot,j) \tilde{S}^\intercal Y_{t-1}, \sigma_j^2) 
\end{equation}
The variance is endowed with a Jeffery prior $\pi(\sigma_j^2) \propto \sigma_j^{-2} \ \  \forall j \in [p]$

Now, for the modelling of $R$, we do it column-wise, in particular, $ \forall j \in [p] $
\begin{align}
	R(\cdot,j) | \gamma_j &\sim  SSL(\gamma_j, \lambda_{0,j}, \lambda_{1}), \\
	 \gamma_j |\theta_j &\sim Bernoulli(\theta_j), \\ 
	  \theta_j  &\sim {\mathcal B} (p-j+1,p)  
\end{align}

The essence of the model is that $R$ is modelled column-by-column with $\gamma_j$ mixxing of different spikes $\lambda_{0,j}$. The proportion is governed by $\theta_j$, which in turn is drawn from a Beta distribution from levelling priors. We consider modelling parameter to be $R, \Sigma$, and the estimator takes the standard maximum-a-posterior approach, namely \begin{equation}
(\hat{R}, \hat{\Sigma}) = \argmax_{R,\Sigma} \pi(R,\Sigma | Y_0,..., Y_T) 
\label{BE}
\end{equation}
where $\pi(\cdot,\cdot | Y_0,..., Y_T)$ is the posterior probability after conditional marginalisation on $\gamma_j$ and likelihood updates as per \autoref{Gaussian}. 

Subsequently, rank can be taken as $\hat{r} = rank(\hat{R})$.

\subsection{Algorithm: a brief overview}
The solution to \autoref{BE} can be approximated through EM iterations column-wise (see \cite{Yang2023, RRG2021} for further discussions). We give a brief overview on the algorithms employed in this paper.

	\begin{algorithm}
	\caption{Overview of column-wise SSL through EM}
	\label{alg}
	\begin{flushleft}
		\textbf{Input}: data $Y_0,..., Y_T$, initialisers $\theta^{(0)}, R^{(0)}, (\sigma_j^{(0)})_{j\in[p]}, $ and initialised SSL parameter $ \lambda_1$, seed for Uniform draws	\\ 
		\textbf{Output}: estimated ranks $ \hat{r}$ and matrix $\hat{R}$
	\end{flushleft}
	\begin{algorithmic}[1]
		\STATE Initialise estimate $\hat{r}=p$, pre-process data (centralisation of $\Delta Y_t$ and normalisation of $\tilde{S}^\intercal Y_t$). Initialise $\lambda_{0,j} = \lambda_1 \ \forall j \in [p]$
		\WHILE{$\hat{r}\geq \frac{T}{\sqrt{p}}$}
		\STATE Increase $\lambda_{0,j}$ by the same amount for all $j$
		\STATE Compute new $\hat{R}$ and $\hat{r}$
		\ENDWHILE
		\STATE Initialise set $P=\{j: \hat{R}(\cdot,j) \neq  \boldsymbol{0} \}$. Set seed for sampling.

		\WHILE{$\hat{r}$ continues to decrease}
		\STATE Sample $j \sim {\mathcal U}(P)$
		\STATE Increase $\lambda_{0,j}$ 
		\STATE Compute new $\hat{R}(\cdot,j)$ 
		\IF{$\hat{\beta_j} = \boldsymbol{0}_p$}
		\STATE	Update $P$ by $P \setminus \{j\}$ and $\hat{r}$ by $\hat{r}-1$
		\ENDIF 
		\ENDWHILE
		\RETURN estimated rank  $\hat{r}$ with the final matrix $\hat{R}$
	\end{algorithmic}
\end{algorithm}

The algorithm \autoref{alg} offers a pragmatic and efficient approach in dealing with finding suitable $\lambda_{0,j}$ and the corresponding sparsity estimation. In particular, starting from line 2, a desired sparsity is declared, pushing $\lambda_{0,j}$ to increase evenly\footnote{Evenly in the sense that $\forall j, j^\prime, \lambda_{0,j} = \lambda_{0, j^\prime}$ so the entire $\lambda_0$ vector has no variation in its element.} until the mild condition of identifiability is reached --- here we put $\hat{r}< \frac{T}{\sqrt{p}}$ to facilitate possible ranks of 10 to 20 when $p \approx 1000$, with $T\approx 500$. Followed by the initial search, a non-zero set of columns are established (in line 6), with further stochastic increase of $\lambda_{0,j}$ being progressed. This makes the algorithm practical, and allows probabilistic inference --- as each seed may result in different magnitude of $\lambda_{0,j}$. Practically, however, with small increment, the resulting $\hat{r}$ often concentrates to the same value.

\section{Stationary Portfolios Through Cointegration}\label{SPC}

\begin{table*}[h]
	\centering
	\begin{tabular}{lrrrrrrrrrrrrrr}
		\toprule
		End of training period	& 22-12 & 23-01 & 23-02 & 23-03 & 23-04 & 23-05 & 23-06 & 23-07 & 23-08 & 23-09 & 23-10 & 23-11 & 23-12 & 24-01  \\
		\midrule
		Number of Portfolios & 9 & 23 & 6 & 19 & 6 & 17 & 12 & 8 & 10 & 9 & 11 & 11 & 14 & 13 \\
		Mean volatility &2.1 & 1.9 & 0.7 & 0.7 & 1.0 & 0.8 & 0.9 & 0.9 & 6.4 & 1.5 & 5.3 & 1.9 & 4.4 & 3.5 \\
		Median volatility &  0.7 & 0.7 & 0.8 & 0.7 & 0.7 & 0.7 & 0.9 & 0.9 & 8.7 & 0.8 & 0.8 & 0.9 & 0.9 & 1.1 \\
		US Benchmark &21.2 & 21.2 & 21.2 & 21.3 & 21.3 & 21.3 & 21.4 & 21.3 & 21.4 & 21.4 & 21.4 & 21.4 & 21.1 & 20.8 \\
		European Benchmark & 17.3 & 17.3 & 17.3 & 17.7 & 17.7 & 17.6 & 17.6 & 17.6 & 17.6 & 17.5 & 17.6 & 17.3 & 17.1 & 16.6 \\
		\bottomrule
	\end{tabular}
	\caption{Summary of portfolio volatility (in percentage) in the in-sample training periods}
	\label{3A}
\end{table*}

\begin{table*}[h]
	\centering
	\begin{tabular}{lrrrrrrrrrrrrrr}
		\toprule
		Testing period& 23-01 & 23-02 & 23-03 & 23-04 & 23-05 & 23-06 & 23-07 & 23-08 & 23-09 & 23-10 & 23-11 & 23-12 & 24-01 & 24-02 \\
		\midrule
		Number of Portfolios & 9 & 23 & 6 & 19 & 6 & 17 & 12 & 8 & 10 & 9 & 11 & 11 & 14 & 13 \\
		Mean volatility & 2.3 & 2.0 & 1.0 & 1.1 & 1.3 & 0.9 & 1.4 & 1.0 & 3.7 & 1.7 & 4.0 & 1.8 & 3.0 & 2.6 \\
		Median volatility &1.2 & 1.3 & 1.0 & 1.1 & 1.2 & 0.9 & 1.4 & 1.0 & 4.7 & 1.3 & 1.1 & 1.1 & 1.2 & 1.3 \\ 
		US Benchmark & 16.9 & 15.9 & 19.5 & 13.1 & 14.4 & 12.3 & 9.7 & 14.8 & 12.9 & 16.8 & 12.6 & 12.4 & 14.5 & 15.5 \\
		European Benchmark & 11.0 & 11.7 & 21.8 & 8.7 & 13.0 & 10.9 & 17.1 & 10.3 & 11.6 & 13.8 & 11.2 & 6.1 & 11.2 & 8.6 \\
		\bottomrule
	\end{tabular}
	\caption{Summary of portfolio volatility (in percentage) in the out-of-sample testing periods}
	\label{3B}
\end{table*}
\subsection{Portfolio construction upon cointegration}
From \autoref{Method}, we obtain $\hat{R}$ and therefore $\hat{\Pi}$ as mentioned in $\autoref{hatPi}$. Due to the orthogonality of $\tilde{S}$, we would have $\hat{r}$ many linearly independent non-zero rows in $\hat{\Pi}$. The idea then, is to use these rows to construct portfolios.

Let $\hat{\Pi}_i$ be a non-zero row of $\hat{\Pi}$, define a stationary portfolio as $\alpha_i:= \frac{\hat{\Pi}_i}{||\hat{\Pi}_i||_1}$. The interpretation of $\alpha_i$ is the weight allocated to each stock in a long-short portfolio, summed to one. Additionally, $\hat{r}$ is interpreted as the number of stationary portfolios.

\subsection{Data usage}
In this paper, we present estimation results based on three asset spaces, on a rolling training \& testing basis. Standard ADF tests are run to test whether the individual time series are $I(1)$.

The first asset space is S\&P 500 index (SPY) constituents, with the index being considered as a benchmark for the US market. The list is obatined from the index-tracking ETF with ticker IVV, dated end of year 2023. There are 453 stocks which are tested to be $I(1)$. The second asset space is STOXX Europe 600 (SXXP) index constiuents, with the index being the benchmark for the European market. The list is obatined from the index-tracking ETF with ticker EXSA, also dated end of year 2023. There are 544 stocks which are tested to be $I(1)$. The third space is a combination of the first and the second, where the aforementioned US and European markets are combined together, resulting in 997 stocks.

In terms of the treatment towards number of observations over time, we consider a rolling estimation approach for training and testing of the parameter estimation and induced portfolio. The first training period is from start of year 2021 to end of year 2022, spanning two years, resulting in 517 observations. The subsequent testing period is the January 2023. The proceeding training period is from the start of February 2021 to the end of January 2023, with testing period being the February 2023. The rolling proceeds for 14 times with the testing period ending at the end of February 2024.

To summarise, the feature dimension $p$ is, respectively, 453, 544, and 997 in the first to third space, and $T$ is circa 510 to 520, depending on day-counts in the rolling periods.

\subsection{Summary of Performance}
The virtue of stationary portfolio is in volatility reduction. In \autoref{3A}, we summarise the estimation result in-sample. Over the three estimations, we found a range of 6 to 23 portfolios over time, with remarkably low volatility amongst them, demonstrated both in mean and median across the portfolios, and the reporting number can often be 90\% or more lower compared to US or European benchmarks (the index volatility of SPY and SXXP respectively). 

We proceed to test the same portfolio in testing period. In \autoref{3B}, we illustrate the testing result, where volatilities are computed upon the same portfolios as they were trained. For instance, the first set of 9 portfolios were trained based on data up to December 2022 (first column of \autoref{3A}) --- these are then run in the testing period in January 2023. As reported, the low volatility is persistent throughout the out-of-sample testing period, which means the cointegrated portfolio tends to be robust in its low-volatility property in the testing period proceeding the in-sample training dataset.

\subsection{Sample Portfolio Insight}
\begin{figure}[h]
	\centering
	\includegraphics[width=\linewidth]{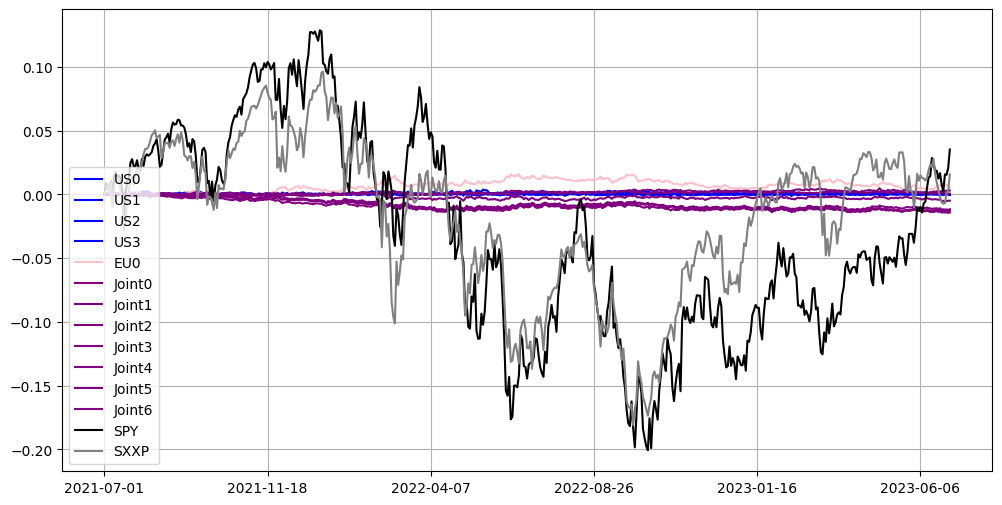}
	\includegraphics[width=\linewidth]{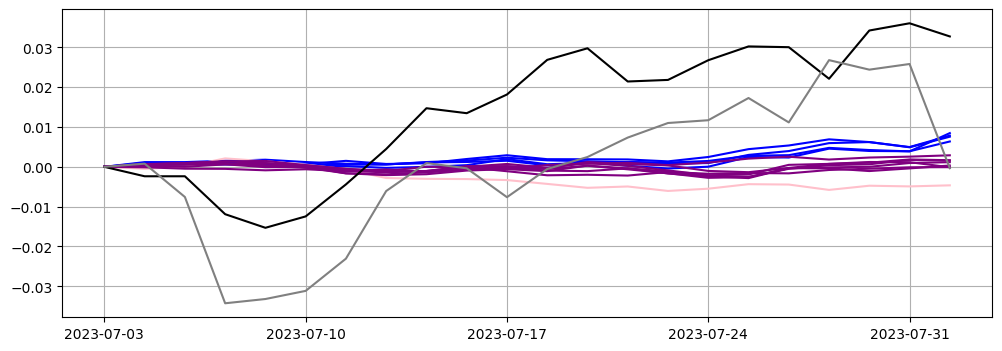}
	\caption{Typical training and testing performance in portfolio returns}
	\label{fig:3a}
\end{figure}
In \autoref{fig:3a}, we plot a typical estimation result to showcase the portfolio performance, where the rolling in-sample training period is July 2021 to June 2023, with testing being July 2023. 12 portfolios are generated from an estimated $\hat{r}=4$ in the US market, $\hat{r}=1$ in the European market, and $\hat{r}=7$ in the joint market. 

In-sample performance saw a clear low-volatility behaviour, which carries over to the testing period, admitting some shifting towards the end --- typical behaviour of financial time series which have potential change-points in the un-seen periods.

\begin{figure}[h]
	\centering
	\includegraphics[width=\linewidth]{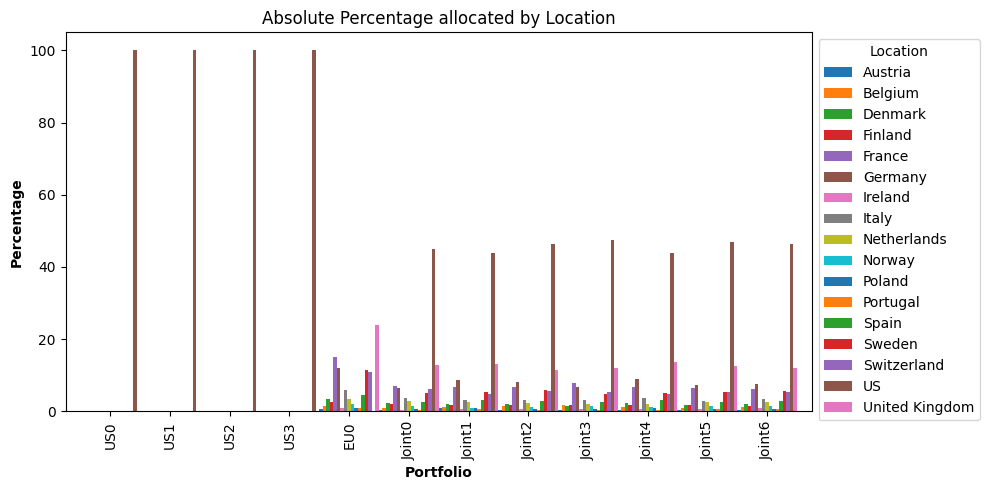}
	\includegraphics[width=\linewidth]{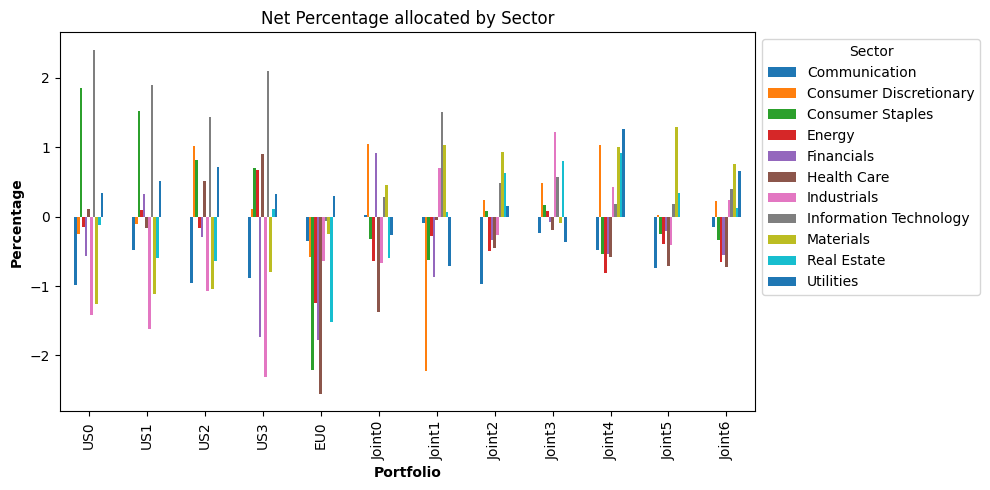}
	\caption{Portfolio analytics: absolute percentage by location (upper) and net percentage by sectors (lower)}
	\label{fig:3b}
\end{figure}

\autoref{fig:3b} is drawn for illustration of insight and portfolio analysis towards the aforementioned performance. Commonality may be observed amongst the estimated portfolios --- for the joint portfolio, summed absolute percentages (grouped $|\alpha_{i,j}|$ for some $j$-th entry in $\alpha_i$) tend to allocate a good portion to the US, with the joint portfolios focusing on the UK as the secondary choice in geography. Net percentages (grouped $\alpha_{i,j}$, hence some positive and some negative) depict differences across each portfolio, where some (mainly when the US market space is individually estimated) have net positive on Information Technology, whilst others are more settled.

\section{Combining Stationary Portfolio to Improve Sharpe Ratio}\label{SPcombo}

\begin{table*}[h]
	\centering
	\begin{tabular}{lrrrrrrrrrrrrrr}
		\toprule
		End of training period	& 22-12 & 23-01 & 23-02 & 23-03 & 23-04 & 23-05 & 23-06 & 23-07 & 23-08 & 23-09 & 23-10 & 23-11 & 23-12 & 24-01  \\
		\midrule
		Number of Portfolios & 2 & 9 & 3 & 2 & 3 & 2 & 4 & 1 & 5 & 2 & 4 & 5 & 3 & 2 \\
		SR Optimised & 1.3 & 2.6 & 0.6 & 0.8 & -0.2 & -0.2 & 1.1 & 2.0 & 0.4 & 0.3 & -2.6 & 1.2 & -0.3 & 2.6 \\
		US Benchmark & 1.9 & 4.1 & 0.9 & 1.2 & -0.3 & -0.3 & 1.8 & 2.7 & -0.2 & -0.9 & -5.7 & 0.7 & -0.4 & 4.0 \\
		\bottomrule
	\end{tabular}
	\caption{Summary of annualised portfolio returns (in percentage) in the in-sample training periods(in the SPY only case)}
	\label{4A}
\end{table*}

\begin{table*}[h]
	\centering
	\begin{tabular}{lrrrrrrrrrrrrrr}
		\toprule
		Testing period& 23-01 & 23-02 & 23-03 & 23-04 & 23-05 & 23-06 & 23-07 & 23-08 & 23-09 & 23-10 & 23-11 & 23-12 & 24-01 & 24-02 \\
		\midrule
		SR Optimised & 56.0 & -29.8 & 32.2 & 8.5 & 2.8 & 51.5 & 28.1 & -13.6 & -22.2 & -16.2 & 54.2 & 34.6 & 26.8 & 35.4 \\
		US Benchmark & 81.9 & -48.3 & 51.2 & 14.6 & 3.9 & 74.4 & 43.2 & -22.4 & -73.8 & -30.7 & 107.0 & 56.8 & 33.3 & 61.6 \\
		\bottomrule
	\end{tabular}
	\caption{Summary of annualised portfolio returns (in percentage) in the out-of-sample training periods (in the SPY only case)}
	\label{4B}
\end{table*}
\subsection{Motivation and Methodology}

Let $\mu(\alpha)_t, \sigma(\alpha)_t$ be the daily return and (annualised) volatility of a portfolio $\alpha$ over day $t$. The Sharpe ratio is defined as $SR(\alpha) = \mE[\frac{\mu(\alpha)_t}{\sigma(\alpha)_t}]$, and empirically the expectation is evaluated over the corresponding data. Given the stationary portfolio and its low volatility property we have demonstrated, it is tempting to pose the following question: can a combination of low volatility cointegrated portfolios and a passive strategy yield better Sharpe ratio?  Intuitively, this may be seen as a classical notion of risk-return trade-off (back to the Markowitz portfolio selection), where Sharpe ratio may be improved by a reduction in return and volatility, due to the mixing of stationary portfolio.

To consider a combination portfolio, we first consider a simple setting to build up the financial intuition. Let vector $\alpha$ be such that $\alpha Y \sim I(0)$ and $||\alpha||_1=1$, which represents a typical cointegrated portfolio depicting stationarity. Let $\beta$ be a passive investment strategy (SPY or SXXP in this paper), it's likely that $\beta Y \sim I(1)$ and empirically yielding higher volatility. The combination strategy concerns a new portfolio $\theta \in (0,1)$ where the resulting return takes the form of $\Psi= \theta \alpha Y + (1-\theta) \beta Y$. Intuitively, while $\Psi$ could have lower return due to its allocation into cointegrated portfolio, it also could have lower volatility --- perhaps lower to the extent that the Sharpe ratio becomes higher. 

Certainly, as shown empirically, we can often find multiple stationary portfolios over the training data, so the practical optimisation is framed as follows. Given $r$ number of portfolios and $b$ number of baseline passive investment strategies\footnote{ $b$ takes a value of 1 or 2 in this paper, indicating a choice of US only or a choice amongst US and European benchmarks}, written as matrices $\alpha := (\alpha_1,..., \alpha_r)$ and $\beta := (\beta_1, ... ,\beta_b)$ respectively, the combination portfolio takes the form of  $\Psi = \theta_{1:r}^\intercal \alpha Y + \theta_{r+1:r+b}^\intercal \beta Y$ where $\theta \in [-1,1]^{r+b}$ such that $||\theta||_1=1$

In the training dataset indexed by $t\in \texttt{train}$, we opt to optimise the following objective:
\begin{align}
\text{maximise}  \ \	SR(\Psi (\theta)) &:= \mE_{t \in \texttt{train}} \left[\frac{\mu(\Psi(\theta))_t}{\sigma(\Psi(\theta))_t} \right] \label{SRmaximisation}
\\	
\text{subject to}  \ \  \theta & \in \Theta
	\end{align}

The maximiser $\theta^*$ is then taken to construct the optimised portfolio, proceeding into out-of-sample testing, where we aim to answer the initial question: whether $\theta^*$, which combines the $r$ many cointegrated portfolios can improve Sharpe ratio.

The space $\Theta$ is set to such that some passive mixing are required --- as far as this section is concerned, as a mainstream application. Specific definitions are set in each of the cases in the proceeding parts of this section.

\subsection{US only}
We first present the result where asset space is constrained in US. In this setting, $b=1$ and $\Theta$ is set to such that $\theta_{r+1} \in (\frac{1}{2}, 1)$, meaning the combination portfolio must take at least half of the benchmark, which is SPY. Certainly, choosing how much the cointegrated portfolios are allocated to would be up to the optimisation objective.

	\begin{figure}[h]
	\centering
	\includegraphics[width=\linewidth]{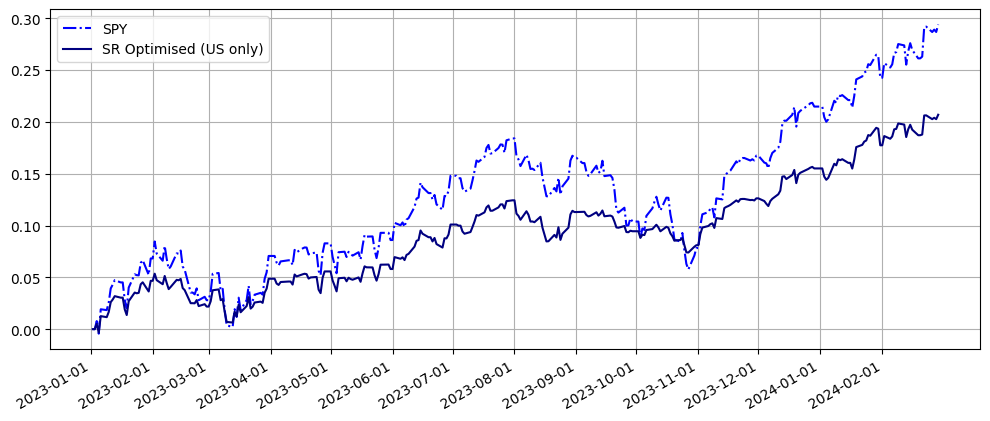}
	\caption{Portfolio returns for US optimised portfolio against its Benchmark in the out of sample period}
	\label{fig:4a}
\end{figure}

In \autoref{4A}, we present the in-sample training period performance and its optimised return, comparing against the benchmark. Total number of portfolios ($r$) in each of the rolling training period is reported in the first row, followed by the annualised return in the period in the second row. There is a general moderation of return, compared to the benchmark, which underlines the nature that stationary portfolio choices ought not to be profitable by themselves, even for the optimised in-sample combination. We proceed into the testing period, which are reported in \autoref{4B} --- there was a significant upward trend in the market in those period, as can be seen in the benchmark. Albeit, the optimised portfolio keeps up the general trend of up-tick while preserving lower negative returns in phases of downturns. In \autoref{fig:4a}, we summarise the cumulative returns in all testing period and plot against the benchmark. The volatility has clearly become lower, with cumulative return also being less than the benchmark.

Out of sample Sharpe ratio of the optimised portfolio for the period is reported at 1.96, while the benchmark is at 1.70. Maximum drawdown of the portfolio is at 4.5\% while the benchmark is at 10.7\%. More analyses of drawdown and risk management are discussed in \autoref{RM}.

\subsection{US and European markets}

\begin{table*}[h]
	\centering
	\begin{tabular}{lrrrrrrrrrrrrrr}
		\toprule
		End of training period	& 22-12 & 23-01 & 23-02 & 23-03 & 23-04 & 23-05 & 23-06 & 23-07 & 23-08 & 23-09 & 23-10 & 23-11 & 23-12 & 24-01  \\
		\midrule
		US Portfolios & 2 & 9 & 3 & 2 & 3 & 2 & 4 & 1 & 2 & 2 & 4 & 5 & 3 & 2 \\
		European Portfolios & 4 & 2 & 1 & 1 & 1 & 1 & 1 & 1 & 2 & 2 & 1 & 3 & 5 & 7 \\
		Joint Portfolios & 2 & 10 & 2 & 16 & 2 & 14 & 7 & 6 & 1 & 5 & 3 & 3 & 3 & 2 \\
		SR Optimised I & 2.0 & 4.5 & 3.3 & 1.6 & 1.7 & 0.2 & 1.0 & 1.1 & -0.4 & 0.5 & -4.2 & 0.7 & -0.5 & 2.0 \\
		SR Optimised II & 2.0 & 4.4 & 3.3 & 1.6 & 1.6 & 0.2 & 1.0 & 1.2 & -0.7 & -0.2 & -4.2 & 0.2 & -0.3 & 2.4 \\
		
		European Benchmark & 2.9 & 6.5 & 6.1 & 3.2 & 3.3 & 0.2 & 0.8 & 0.9 & -1.9 & -0.3 & -5.6 & -1.2 & -1.4 & 1.3  \\
		\toprule
		Testing period& 23-01 & 23-02 & 23-03 & 23-04 & 23-05 & 23-06 & 23-07 & 23-08 & 23-09 & 23-10 & 23-11 & 23-12 & 24-01 & 24-02 \\
		\midrule
		SR Optimised I & 47.3 & -1.6 & 12.6 & 14.0 & -14.5 & 24.7 & 17.5 & -5.5 & -38.8 & -17.4 & 72.3 & 30.0 & 23.5 & 24.4 \\
		SR Optimised II & 48.6 & 0.3 & 14.5 & 14.0 & -13.2 & 25.0 & 17.6 & -8.5 & -41.7 & -17.7 & 71.9 & 39.3 & 13.3 & 38.4 \\
		European Benchmark  & 56.8 & 24.0 & 0.5 & 26.6 & -27.8 & 19.9 & 31.0 & -7.9 & -23.6 & -35.6 & 74.8 & 38.3 & 33.7 & 32.1 \\
		\bottomrule
	\end{tabular}
	Remark:	SR Optimised I refers to the portfolio optimised using all 3 types (US, European and Joint) whereas SR Optimised II refers to the one optimised using only Joint Portfolios.
	\caption{Summary of annualised portfolio returns (in percentage) in the in-sample training period and out-of-sample testing period (in the joint asset space)}
	
	\label{4C}
\end{table*}

	\begin{figure}[h]
	\centering
	\includegraphics[width=\linewidth]{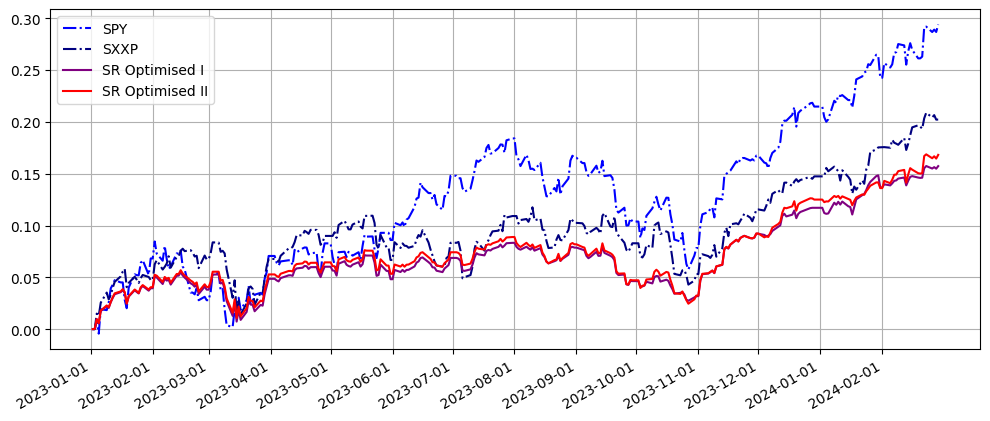}
	\caption{Portfolio returns for jointly optimised portfolio against its Benchmark in the out of sample period}
	\label{fig:4b}
\end{figure}

We now enlarge the asset space to include European markets. In this setting, $b=2$, and $\Theta $ is imposed to be such that $\theta_{r+1}, \theta_{r+2} \in (\frac{1}{4},\frac{1}{2})$. This is a moderated version to the US only case, wherein the benchmark composition should be at least a quarter each (totalled to half, which is the restriction in the US only case), while not allowing any one to fully dominate the portfolio.

There are two ways to optimise such a combination of two markets. The first being taking all three sets of estimations (the one in US market, the one in European market, and the one in the joint estimation) and treat all portfolios into the same optimisation control --- this is annotated as "SR Optimised I" in \autoref{4C} and \autoref{fig:4b}. In this case, the size of $r$ varies from 6 to 21, with detailed compositions being reported in the first three rows in \autoref{4C}. The other way is to take the joint estimations only (i.e. the joint portfolios as reported in \autoref{4C}), as arguably the joint estimations capture the entire dataset already, albeit the statistical risk of modelling a vastly sparse model (recall that the features in joint asset space being more than twice as much as the US only case, and that the $\Pi$ matrix would then make the dimension to be more than four times as much). In this case, $r$ varies from 1 to 16, and certainly strictly less than the first way of combination, as the candidate portfolios are a subset of the first way's. This method is annotated as "SR Optimised II" in the reporting results.

Both optimised portfolios depict a similar trend as was observed in the US only case, wherein they capture the general trend while having less volatility and drawdown. The difference between the two SR optimised portfolios are rather small --- this means the US-only or European-only portfolios are relatively redundant when it comes to the optimal combination into Sharpe ratio optimisation.

Out of sample Sharpe ratio of the portfolios are 1.76 and 1.80 for I and II respectively, comparing to a benchmark of 1.7 and 1.42 for the US and European indices. Max drawdown for the two portfolios are reported at 5.2\% and 5.9\% respectively, whereas the ones for US and European indices are 10.7\%  and 6.7\%. 

In this subsection, we may nominally observe an interesting phenomenon wherein extension into a wider European market does not further improve the Sharpe ratio and Max Drawdown compared to the US market alone. However, this may be explained by general poorer performance compare to the US market, as the index does have a weaker Sharpe ratio by itself, so the extension may be understood as a cost of diversification in the relevant period. Furthermore, the optimised portfolio still outperforms both indices despite having to include at least $\frac{1}{4}$ of the European index, underlining the strength of such optimised portfolio.

\section{Further Application to Risk Management}\label{RM}

\subsection{Drawdown reduction}
There are no mentioning of pure volatility minimisation or drawdown reduction in the objective function as per \autoref{SRmaximisation}, however, what can be explicitly observed from the result is that not only Sharpe ratio, in out-of-sample, outperforms the benchmarks, so do the risk metrics --- here we focus on the drawdown.
\begin{figure}[h]
	\centering
	\includegraphics[width=\linewidth]{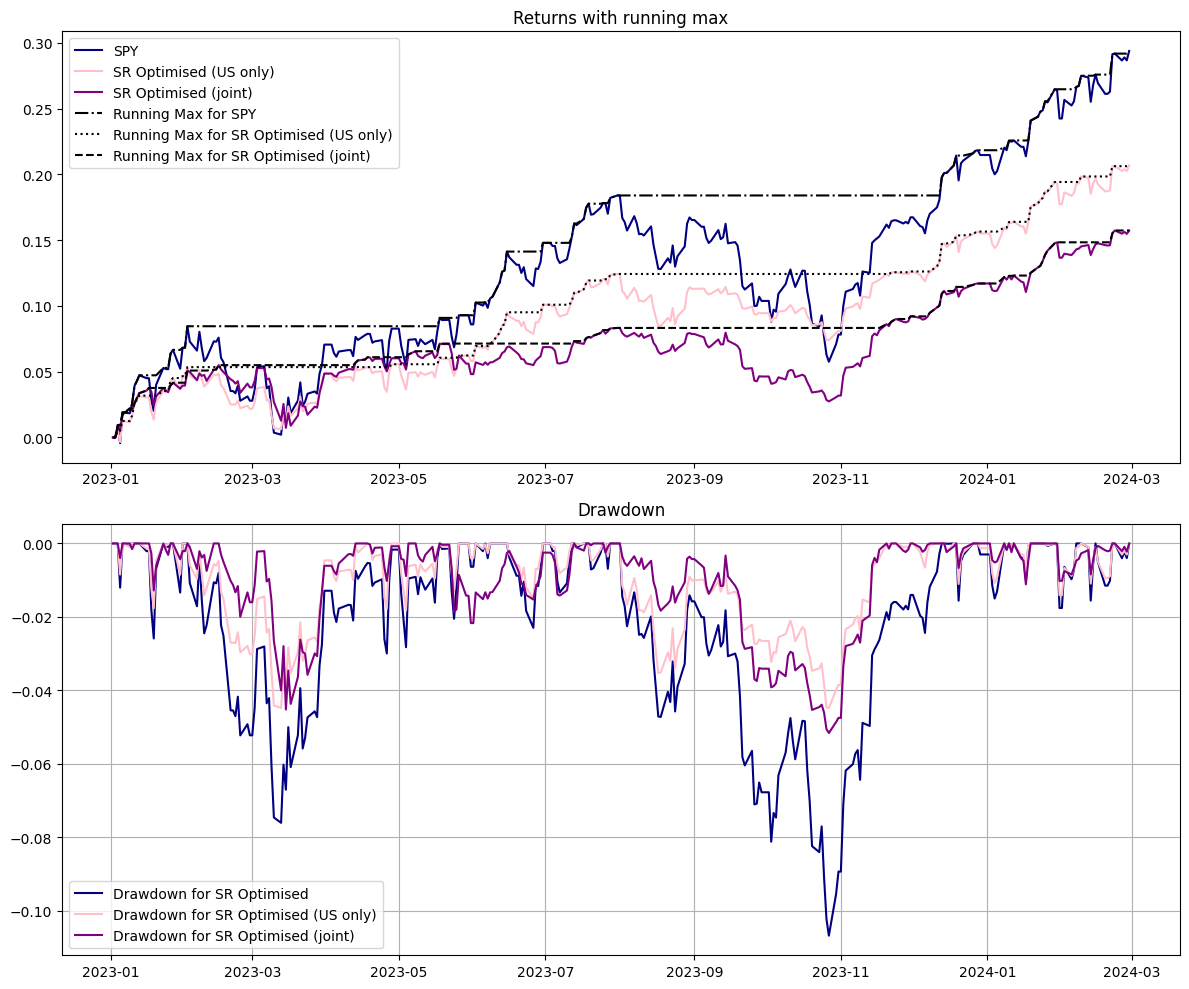}
	\caption{Further illustration of drawdown: US and joint optimised portfolio against US benchmark}
	\label{fig:5a}
\end{figure}

In \autoref{fig:5a}, we present the cumulative returns of SPY and the two SR optimised portfolios, one in the US-only asset space and another as the SR Optimised I case in the joint asset space. The running drawdown can be seen in the bottom plot, induced from the running max in the upper plot.

From the running drawdown, a clear indication can be drawn to the benefit of low-volatility combination with cointegrated portfolio, wherein the optimised portfolios managed to half the drawdown in many running cases, and achieving no significant idiosyncratic downside.

\begin{table*}[h]
	\centering
	\begin{tabular}{lrrrrrrrrrrrrrr}
		\toprule
		End of training period	& 22-12 & 23-01 & 23-02 & 23-03 & 23-04 & 23-05 & 23-06 & 23-07 & 23-08 & 23-09 & 23-10 & 23-11 & 23-12 & 24-01  \\		\midrule
		Mean volatility  &&&&&&&&&&&&&&\\of candidate portfolios & 2.1 & 1.9 & 0.7 & 0.7 & 1.0 & 0.8 & 0.9 & 0.9 & 6.4 & 1.5 & 5.3 & 1.9 & 4.4 & 3.5 \\
		Minimised Volatility & 0.0 & 0.0 & 0.0 & 0.0 & 0.0 & 0.0 & 0.0 & 0.0 & 0.0 & 0.0 & 0.0 & 0.0 & 0.0 & 0.0
		\\
		Average Benchmark & 19.3 & 19.3 & 19.2 & 19.5 & 19.5 & 19.4 & 19.5 & 19.4 & 19.5 & 19.4 & 19.5 & 19.4 & 19.1 & 18.7 \\
		\toprule				
		Testing period& 23-01 & 23-02 & 23-03 & 23-04 & 23-05 & 23-06 & 23-07 & 23-08 & 23-09 & 23-10 & 23-11 & 23-12 & 24-01 & 24-02 \\
		\midrule
		Mean volatility  &&&&&&&&&&&&&&\\of candidate portfolios & 2.3 & 2.0 & 1.0 & 1.1 & 1.3 & 0.9 & 1.4 & 1.0 & 3.7 & 1.7 & 4.0 & 1.8 & 3.0 & 2.6 \\
		Minimised Volatility & 0.5 & 0.3 & 0.5 & 0.2 & 0.4 & 0.2 & 0.2 & 0.3 & 0.3 & 0.9 & 0.3 & 0.2 & 0.5 & 0.3 
		\\
		Average Benchmark  & 14.1 & 13.9 & 20.7 & 10.2 & 13.5 & 11.6 & 12.2 & 12.8 & 12.2 & 15.2 & 12.0 & 9.3 & 12.6 & 12.1 \\
		\bottomrule
	\end{tabular}
	\caption{Summary of portfolio volatility (in percentage) in the in-sample and out-of-sample testing periods, with joint benchmark being the average of US and European benchmarks}
	\label{5A}
\end{table*}

\subsection{Volatility minimisation}
In the preceding discussions, we focused on \autoref{SRmaximisation} as the objective. Here, instead, we trial an alternative objective where we only focus on volatility minimisation. Particularly, consider \begin{equation}
\text{minimise}  \ \	 \mE_{t \in \texttt{train}} \left[{\sigma(\Psi(\theta))_t} \right] 
\end{equation}
instead of \autoref{SRmaximisation}. We do this to both the US-only asset space and joint asset space, with the $\Theta$ space relaxed into $\theta_{r+1:r+b} \in (0,1)^b$ instead of having a minimum holding requirement --- demanding less passive holdings to see how far the cointegrated portfolios stretch to the volatility reduction.

In \autoref{5A}, results are presented in both the training and testing periods --- the training volatility is brought down to below 0.01\% from combining the $r$ many candidate portfolios. Though this is not preserved to testing, the same portfolio still yields more than half of the mean volatility in the candidate portfolios, which are already extremely low compared to the Benchmark.

\begin{figure}[h]
	\centering
	\includegraphics[width=\linewidth]{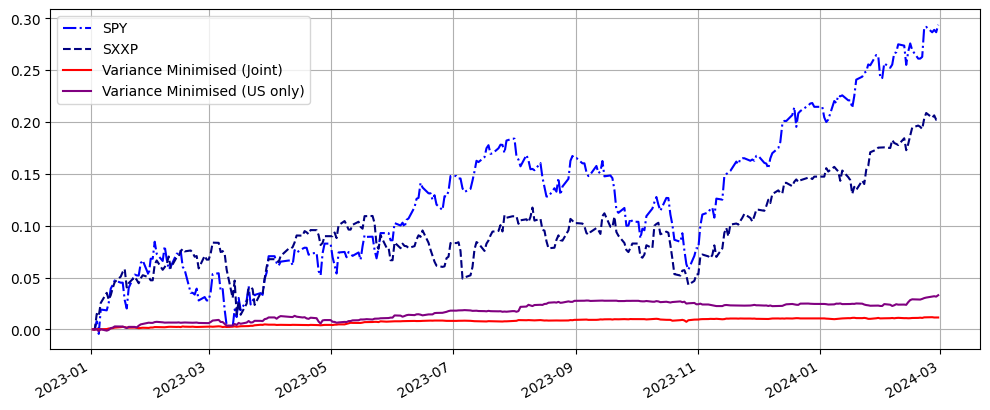}
	\caption{Volatility minimisation: US and jointly optimised portfolio against benchmarks}
	\label{fig:5b}
\end{figure}

We concatenate all testing periods and summarise the performance in \autoref{fig:5b}. Visually, the return can be seen as markedly smooth and drastically less volatile than benchmarks. The Sharpe ratio for US only is obtained at 2.26 and for the joint one being 2.40, while maximum drawdown are 0.7\% and 0.3\% respectively, which are more than 90\% cut from the benchmarks.

\section{Discussions and Conclusion}
\subsection{Summary of results}

To summarise various portfolio findings in this paper, we first recall \autoref{SPC} for the reported volatility in stationary portfolios --- the volatility amongst most of the portfolios are markedly lower than benchmark, which also reflect in a flat cumulative return. This motivates investigations in \autoref{SPcombo} to mix the cointegrated portfolio with some elements of passive strategies. A summary of all the Sharpe ratio and Max Drawdown reported in the out-of-sample periods is presented in \autoref{Summarytab}, as well as the extended investigation into pure volatility minimisation exercise in \autoref{RM}.

\begin{table}[h]
	\begin{tabular}{l|rr}
		\toprule
		&	Sharpe ratio & Max Drawdown \\  \midrule
		\autoref{SPcombo} US only& 1.96& 4.5\% \\ 
		\autoref{SPcombo} US and European & 1.76& 5.2\% \\ 
		\autoref{SPcombo} Joint only& 1.80& 5.9\% \\ 
		\autoref{RM} US only& 2.26& 0.7\% \\ 
		\autoref{RM} US and European & 2.4& 0.3\% \\ 
		US Benchmark &  1.70 & 10.7 \%  \\
		European Benchmark &  1.42 & 6.7 \%  \\
		\bottomrule
	\end{tabular}
	\caption{Summary of Sharpe ratio and Maximum Drawdown in out-of-sample testing periods}
	\label{Summarytab}
\end{table}

The inclusion of cointegrated portfolios has induced lowered return, but matched with much lower volatility --- as they imply a higher Sharpe ratio in various settings of asset space. Drawdown analysis is further extended to showcase the cointegrated portfolios as potential instruments for risk management. We further extend this into a venue of pure volatility minimisation exercise and find both Sharpe ratio and Max Drawdown to be reported at best amongst all.

\subsection{Methodological extension}
Ongoing extensions are investigated on the methodological side, both in the models and in algorithms. For instance, one may criticise the model (as per \autoref{Method}) to be not fully Bayesian, as it rests on a partial least square pre-estimation. Though the rationale has been that it reduces the complexity of rank determination into sparsity of the $R$ matrix, we are in search of more Bayesian approaches to fully express the model in Bayesian methods. This may involve other literature such as \cite{XD2023} for high dimensional matrix estimation.
\subsection{More financial applications}
There is admittedly less attention to potential practical problems arising from the cointegrated portfolios, such as foreign exchange rates (we purely took the returns for all assets, albeit the European stocks are quoted in local currencies), transaction costs, and adjustments of assets in the relevant asset space. Nonetheless, these should not affect the main result as we investigate the stocks with high market capitalisation and that the indices only have minor adjustments of components per quarter. Additionally, as the re-estimation hence rebalancing are done on a monthly basis, the risk and cost of over-trading is mitigated from the start. Therefore, these extensions are relevant, but more towards the implementations by practitioners.

\bibliographystyle{ACM-Reference-Format}
\bibliography{bibliography}
\end{document}